\documentclass[11pt,a4paper]{JHEP3}
\usepackage{amssymb} 
\usepackage{euscript}

\def\a{{\alpha}}
\def\b{{\beta}}
\def\d{\partial}

\def\ket#1{|#1 \rangle}
\newcommand\0{\nonumber}
\newcommand\ES{\EuScript{S}}
\newcommand\one{\mathbb{I}}
\newcommand\T{\EuScript{T}}
\newcommand\X{\EuScript{X}}
\newcommand\V{\EuScript{V}}
\newcommand\U{\EuScript{U}}
\newcommand\R{\EuScript{R}}
\newcommand\ee{\end{eqnarray}}	 	%eqnarray
\newcommand\be{\begin{eqnarray}}
\newcommand\ba{\begin{array}}			%array
\newcommand\ea{\end{array}}
\newcommand\eeq{\end{equation}}	 	%eqnarray
\newcommand\beq{\begin{equation}}

\preprint{SISSA/1/02/EP\\\tt hep-th/0201060}

\title{B field and squeezed states in Vacuum String Field Theory}

\author{ L.Bonora, D.Mamone, M.Salizzoni\\
International School for Advanced Studies (SISSA/ISAS)\\
Via Beirut 2--4, 34014 Trieste, Italy, and INFN, Sezione di 
Trieste\\
E-mail:   \email{bonora@sissa.it}, \email{mamone@sissa.it}, 
\email{sali@sissa.it} }

\abstract{We show that squeezed state solutions for solitonic
lumps in Vacuum String Field Theory still exist in the 
presence of a constant $B$ field. We show in particular that, 
just as in the $B=0$ case, we can write down a compact explicit 
form for such solutions.}


\begin{document}
\section{Introduction}

What happens in Witten's Open String Field Theory, \cite{W1}, when 
a constant $B$ field is switched on? This question has already been 
partially answered. In ref. \cite{W2} and \cite{Sch} it was shown 
that in the limit of field theory the string field theory 
$*$ product factorizes into the ordinary Witten $*$ product and 
the Moyal product. A related result can be obtained in the 
following way. The string field theory action
\be
{\cal S}(\Psi)= - \frac 1{g_0^2} \left[\frac 12 \langle\Psi, Q\Psi\rangle + 
\frac 13 \langle\Psi, \Psi *\Psi\rangle\right]\label{sftaction}
\ee
can be explicitly calculated in terms of local fields, provided 
the string field is expressed itself in terms of local fields 
\be
\ket{\Psi} = (\phi(x)+ A_\mu (x) a_1^{\mu\dagger}+ 
\ldots) c_1\ket{0}\label{stringfield}
\ee
Of course this makes sense in the limit in which string theory can be 
approximated by a local field theory. In this framework 
(\ref{sftaction}) takes the form of an integrated
function $F$ of (an infinite series of) local polynomials 
(kinetic and potential terms) 
of the fields involved in (\ref{stringfield}):
\be
{\cal S}(\Psi) = \int d^{26}x F(\varphi_i,\d\varphi_i,...)
\label{effaction}
\ee
Now, it has been proven by \cite{sugino,KT} that, when a $B$ 
field is switched on, the kinetic term of (\ref{sftaction}) remains 
the same while the three string vertex changes, being multiplied by 
a (cyclically invariant) noncommutative phase factor 
(see (\cite{sugino,KT}) and eq.(\ref{EtoE'})  below). It is easy to 
see on a  general
 basis that the overall effect of such 
noncommutative factor is to replace the ordinary product with the 
Moyal product in the RHS of the effective action
 (\ref{effaction}) \footnote{
Actually in (\cite{sugino,KT}) it was suggested, more drastically, 
that by means of a 
 string field
redefinition one can reduce the modified SFT action to the original 
form (\ref{sftaction}). However it is not clear, at least to us, 
what are the allowed string field redefinitions in SFT. For instance 
in the field theory limit, we have just seen that the terms of the
ordinary effective action are replaced by analogous terms with the 
ordinary product replaced by the Moyal product. But, for instance, 
there is no local field theory redefinition that allows us to pass 
 from such a term as $\int \phi\star\phi\star\phi$, where $\star$ 
denotes the Moyal product, to $\int\phi(x)^3$.
We therefore believe that a closer look at the effects of turning 
on a $B$ field is necessary.}. 

Therefore, we know pretty well the effects of a $B$ field in the 
field theory limit of SFT. What we want to explore in
this paper are the effects of a $B$ field in a nonperturbative 
regime. The good news that comes from the present investigation is that
 a thorough analysis 
of such effects can be carried out for squeezed states and exact 
solutions can be written down for tachyonic lumps.

In the context of the Vacuum String Field Theory (VSFT) squeezed 
states are very important solutions of the SFT equations of motion,
\cite{KP}. There are strong arguments in favor of their
interpretation as D--branes, \cite{RSZ1,RSZ2,RSZ3}, 
see also \cite{David,MT,Mukho}.
For a detailed definition of VSFT see \cite{RSZ1}. The VSFT action 
has the same form as (\ref{sftaction}) with the BRST operator $Q$ 
replaced by a new one, usually denoted ${\cal Q}$, with the 
characteristic of being universal. As a matter of fact 
in \cite{GRSZ}, see also \cite{HK,Oku1,Oku2,RSZ4,RSZ5,Kishi} and
\cite{RSZ3,GT,KO,Moeller},
an explicit representation of ${\cal Q}$ has 
been proposed, purely 
in terms of ghost fields. Now, the equation of motion of VSFT is
\be
{\cal Q} \Psi = - \Psi * \Psi\label{EOM}
\ee
and nonperturbative solutions are looked for in the form
\be
\Psi= \Psi_m \otimes \Psi_g\label{ans}
\ee
where $\Psi_g$ and $\Psi_m$ depend purely on ghost and matter 
degrees of freedom, respectively. Then eq.(\ref{EOM}) splits into
\be
 {\cal Q} \Psi_g & = & - \Psi_g * \Psi_g\label{EOMg}\\
\Psi_m & = & \Psi_m * \Psi_m\label{EOMm}
\ee
Eq.(\ref{EOMg}) will not be involved in our analysis since ghosts 
are unaffected by the presence of a $B$ field. Therefore we will 
concentrate on the solutions of (\ref{EOMm}). 

The value of the action for such solutions is given by 
\be
{\cal S}(\Psi) = {\EuScript K} \langle\Psi_m|\Psi_m\rangle\label{action2}
\ee
where ${\EuScript K}$ contains the ghost contribution. The point is
that, as shown in \cite{RSZ3}, ${\EuScript K}$  is infinite unless 
it is suitably regularized.
It has been argued in \cite{RSZ3} that this should be understood as 
a `gauge' freedom in choosing the solutions of (\ref{EOM}), so that 
a coupled solution of (\ref{EOMg}) and (\ref{EOMm}), even if it 
is naively singular in its ghost component, is nevertheless a 
legitimate representative of the corresponding class of solutions.

The aim of this paper is to find a solutions of eq.(\ref{EOMm})
when a constant $B$ field is switched on along some space 
directions. It is rewarding that not only such solutions exist
in the form of squeezed states, but also that they may be cast
in a simple compact form. We can phrase it shortly by saying that
eq.(\ref{EOMm}) is `exactly solvable' even in the presence of a 
constant $B$ field.

The paper is organized as follows. In the next section we write
down explicitly the algebraic form of
the three string vertex in the presence of a constant $B$ field. 
In section 3 we study the properties of the numerical coefficients
(Neumann coefficients) of the new three string vertex. Finally in 
section 4 we find exact solutions of (\ref{EOMm}) in the form of
squeezed states.

\section{The three string vertex in the presence of a constant background B field}

The three string vertex \cite{W1,GJ1,GJ2} of the Open String Field 
Theory is
given
 by
\be
|V_3\rangle= \int d^{26}p_{(1)}d^{26}p_{(2)}d^{26}p_{(3)}
\delta^{26}(p_{(1)}+p_{(2)}+p_{(3)})\,{\rm exp}(-E)\, |0,p\rangle_{123}\label{V3}
\ee
where
\be
E= \sum_{r,s=1}^3\left(\frac 12 \sum_{m,n\geq 1}\eta_{\mu\nu}
a_m^{(r)\mu\dagger}V_{mn}^{rs}
a_n^{(s)\nu\dagger} + \sum_{n\geq 1}\eta_{\mu\nu}p_{(r)}^{\mu}V_{0n}^{rs}
a_n^{(s)\nu\dagger} +\frac 12 \eta_{\mu\nu}p_{(r)}^{\mu}V_{00}^{rs}
p_{(s)}^\nu\right) \label{E}
\ee
Summation over the Lorentz indices $\mu,\nu=0,\ldots,25$ 
is understood and $\eta$ denotes the flat Lorentz metric and
the operators $ a_m^{(r)\mu},a_m^{(r)\mu\dagger}$ denote the non--zero 
modes matter oscillators of the $r$--th string, which satisfy
\be
[a_m^{(r)\mu},a_n^{(s)\nu\dagger}]= 
\eta^{\mu\nu}\delta_{mn}\delta^{rs},
\quad\quad m,n\geq 1 \label{CCR}
\ee
$p_{(r)}$ is the momentum of the $r$--th string and 
$|0,p\rangle_{123}\equiv |p_{(1)}\rangle\otimes |p_{(2)}\rangle\otimes |p_{(3)}\rangle$ is 
the tensor product of the Fock vacuum 
states relative to the three strings. $|p_{(r)}\rangle$ is annihilated by 
the annihilation 
operators $a_m^{(r)\mu}$ and is eigenstate of the momentum operator 
$\hat p_{(r)}^\mu$ 
with eigenvalue $p_{(r)}^\mu$. The normalization is
\be
\langle p_{(r)}|\, p'_{(s)}\rangle = \delta_{rs}\delta^{26}(p+p')\0
\ee
The coefficients $V_{MN}^{rs}$ ($M(N)$ denotes from now on the couple $\{0,m\}$
($\{0,n\}$)) have been computed in \cite{GJ1,GJ2}. We will use them in the notation
of Appendix A and B of \cite{RSZ2}. 

Our first goal is to find the new form of the coefficients $V_{MN}^{rs}$ when a
constant $B$ field is switched on. We start from the simplest case, i.e. when
$B$ is nonvanishing in the two space directions, 
say the $24$--th and $25$--th ones.
Let us denote these directions with the Lorentz indices $\alpha$ and 
$\beta$.
Then, as is well--known \cite{SW,sugino,KT}, in these two direction 
we have a new effective metric $G_{\alpha\beta}$, 
the open string metric, as well as
an effective antisymmetric parameter $\theta_{\alpha\beta}$, given by
\be
G^{\alpha\beta} = \left(\frac {1}{\eta + 2\pi \alpha' B}\,\eta\, 
\frac {1}{\eta - 2\pi \alpha' B}\right)^{\alpha\beta},\quad\quad
\theta^{\alpha\beta} = -(2\pi \a')^2\left(\frac {1}{\eta + 2\pi \alpha' B}\,B\,
\frac {1}{\eta - 2\pi \alpha' B}\right)^{\alpha\beta}\0
\ee
Henceforth we set $\a' =1$. This is in order to conform to
the convention of \cite{RSZ2}, whose results will be compared with 
ours. 

The three string vertex is modified only in the 24-th and 
25-th direction, which, in view of the subsequent D--brane 
interpretation, we call the transverse directions.
We split the three string vertex into the tensor product of the perpendicular 
part and the parallel part
\be
|V_3\rangle = |V_{3,\perp}\rangle \, \otimes\,|V_{3,_\|}\rangle\label{split}
\ee
The parallel part is the same as in the ordinary case and will not be 
re-discussed here. On the contrary we will describe in detail the 
perpendicular part of the vertex. We rewrite the exponent $E$ as 
$E=E_\|+ E_\perp$, according to the above splitting. 
$E_\perp$ will be modified as follows
\be
E_\perp&\to& E'_\perp = \sum_{r,s=1}^3\left(\frac 12 \sum_{m,n\geq 1}
G_{\alpha\beta}a_m^{(r)\alpha\dagger}V_{mn}^{rs}
a_n^{(s)\beta\dagger} + 
\sum_{n\geq 1}G_{\alpha\beta}p_{(r)}^{\alpha}V_{0n}^{rs}
a_n^{(s)\beta\dagger}\right.\0\\
&&\left. \quad\quad +\,\frac 12 G_{\alpha\beta}p_{(r)}^{\alpha}V_{00}^{rs}
p_{(s)}^\beta+ 
\frac i2 \sum_{r<s} p_\alpha^{(r)}\theta^{\alpha\beta}
p_\beta^{(s)}\right)\label{EtoE'}
\ee

Next, as far as the zero modes are concerned, we pass from the 
momentum to the oscillator basis, \cite{GJ1,GJ2}. We define
\be
a_0^{(r)\alpha} = \frac 12 \sqrt b \hat p^{(r)\alpha} 
- i\frac {1}{\sqrt b} \hat x^{(r)\alpha},
\quad\quad
a_0^{(r)\alpha\dagger} = \frac 12 \sqrt b \hat p^{(r)\alpha} + 
i\frac {1}{\sqrt b}\hat x^{(r)\alpha}, \label{osc}
\ee
where $\hat p^{(r)\alpha}, \hat x^{(r)\alpha}$ are the zero momentum 
and position operator of the $r$--th string, and we have kept the 
`gauge' parameter $b$ of ref.\cite{RSZ2} ($b\sim {\a'}$). 
It is understood that $p^{(r)\alpha} = G^{\a\b}p^{(r)}_\b$. We have 
\be
\big[a_0^{(r)\alpha},a_0^{(s)\beta\dagger}\big]= G^{\alpha\beta}\delta^{rs} 
\label{a0a0} 
\ee
Denoting by $|\Omega_{b,\theta}\rangle$ the oscillator vacuum 
(\,$a_0^\alpha|\Omega_{b,\theta}\rangle=0$\,), 
the relation
between the momentum basis and the oscillator basis is defined by
\be
&&|p^{24}\rangle_{123}\otimes|p^{25}\rangle_{123}\equiv |\{p^\alpha\}\rangle_{123} 
=\0\\
&& \nonumber \\
&&\left(\frac b{2\pi\sqrt {{\rm det}G}}\right)^\frac 32 {\rm exp} \left[\sum_{r=1}^3
\left(- \frac b4 p^{(r)}_\alpha G^{\alpha\beta}p^{(r)}_\beta+
\sqrt b  a_0^{(r)\alpha\dagger}p^{(r)}_\a
- \frac 12 a_0^{(r)\alpha\dagger}G_{\alpha\beta}a_0^{(r)\beta\dagger}
\right)\right]|\Omega_{b,\theta}\rangle\0
\ee

Now we insert this equation inside $E'_\perp$ and try to eliminate the
momenta along the perpendicular directions by integrating them out. 
To this end we rewrite $E'_\perp$ in the following way and, for 
simplicity, drop all the labels $\alpha,\beta$ and $r,s$:
\be
E'_\perp = \frac 12\sum_{m,n\geq 1}a_m^\dagger GV_{mn}a_n^\dagger + 
\sum_{n\geq 1}
pV_{0n}a_n^\dagger + \frac 12 p\left[G^{-1}(V_{00}+\frac b2) + 
\frac i2 \theta \epsilon \chi \right]p
-\sqrt b p  a_0^\dagger + 
\frac 12 a_0^\dagger G a_0^\dagger\0
\ee
where we have set $\theta^{\alpha\beta}= 
\epsilon^{\alpha\beta}\theta$ and introduced
the matrices $\epsilon$ with entries 
$\epsilon^{\alpha\beta}$ and $\chi$ with entries
\be
\chi^{rs}= \left(\matrix{0&1&-1\cr -1&0&1\cr 1&-1&0}\right)\label{chi}
\ee
At this point we impose momentum conservation. There are three 
distinct ways to do that and eventually one has to (multiplicatively) 
symmetrize with respect to them. Let us start by setting 
$p_3=-p_1-p_2$ in $E'_\perp$ and obtain an expression of the form
\be
p\, X_{00}\, p + \sum_{N\geq 0} p\, Y_{0N}\, a_N^\dagger+
\sum_{M,N\geq 0}a_M^\dagger\, Z_{MN }\, a_N^\dagger\label{pXp}
\ee
where, in particular, $X_{00}$ is given by
\be
X_{00}^{\alpha\beta,rs}= G^{\alpha\beta}\, (V_{00}+ \frac b2) \,
\eta^{rs}+ i \frac \theta {4} \,\epsilon^{\a\b}\, \epsilon^{rs}\label{X00}
\ee
Here the indices $r,s$ take only the values 1,2, and
\be
\eta = \left(\matrix{1 & 1/2\cr 1/2 & 1}\right),\quad\quad
\epsilon= \left(\matrix{0& 1\cr -1 &0}\right) \label{epsiloneta}
\ee
Now, as usual, we redefine $p$ so as eliminate the linear term in 
(\ref{pXp}). At this point we can easily perform the Gaussian 
integration over $p_{(1)},p_{(2)}$, while the remnant of 
(\ref{pXp}) will be expressed in terms of the inverse of
$X_{00}$:
\be
\left(X_{00}^{-1}\right)^{\a\b,rs}= 
\frac {2A^{-1}}{4a^2+3}\left(\frac 32\,  G^{\a\b}\,
(\eta^{-1})^{rs} -2i \,a\, \hat\epsilon^{\a\b} \,
\epsilon^{rs}\right)\label{X00-1}
\ee
where
\be
A = V_{00}+ \frac b2,  \quad\quad\quad a = 
\frac \theta{4A }\sqrt{{\rm Det}G},
\quad\quad \epsilon^{\a\b} = 
\sqrt{{\rm Det} G}\, \hat\epsilon^{\a\b}\label{definitions}
\ee
Let us use henceforth for the $B$ field the explicit form 
\be
B_{\a\b}= \left(\matrix {0&B\cr -B&0\cr}\right)\label{B}
\ee
so that
\be
{\rm Det G} = \left( 1+ (2 \pi B)^2\right)^2, \quad \quad
\theta \,\sqrt {\rm Det G} = - (2\pi )^2 B,
\quad\quad a= -\frac {\pi^2}A\, B\label{thetaB}
\ee

Now one has to symmetrize with respect to the three possibilities
of imposing the momentum conservation. Remembering the factors due 
to integration over the momenta and collecting the results one 
gets for the three string vertex in the presence of a $B$ field 
\be
|V_3 \rangle' = |V_{3,\perp}\rangle ' \,\otimes\,|V_{3,\|}\rangle \label{split'}
\ee
$|V_{3,\|}\rangle$ is the same as in the ordinary case 
(without $B$ field), while
\be
|V_{3,\perp}\rangle'= K_2\, e^{-E'}|\tilde 0\rangle\label{V3'} 
\ee
with
\be
&&K_2= \frac {\sqrt{2\pi b^3}}{A^2 (4a^2+3)}({\rm Det} G)^{1/4},\label{K2}\\
&&E'= \frac 12 \sum_{r,s=1}^3 \sum_{M,N\geq 0} a_M^{(r)\a\dagger}
\V_{\a\b,MN}^{rs} a_N^{(s)\b\dagger}\label{E'}
\ee
and $|\tilde 0\rangle = |0\rangle \otimes |\Omega_{b,\theta}\rangle$.
The coefficients $\V_{MN}^{\a\b,rs}$ are given by
\be
&&\V_{00}^{\a\b,rs} = G^{\a\b}\delta^{rs}- \frac {2A^{-1}b}{4a^2+3}
\left(G^{\a\b} \phi^{rs} -ia \hat\epsilon^{\a\b}\chi^{rs}\right)
\label{VV00}\\
&&\V_{0n}^{\a\b,rs} = \frac {2A^{-1}\sqrt b}{4a^2+3}\sum_{t=1}^3
\left(G^{\a\b} \phi^{rt} -ia \hat\epsilon^{\a\b}\chi^{rt}\right)
V_{0n}^{ts}\label{VV0n}\\
&&\V_{mn}^{\a\b,rs} = G^{\a\b}V_{mn}^{rs}-
\frac {2A^{-1}}{4a^2+3}\sum_{t,v=1}^3
V_{m0}^{rv}\left(G^{\a\b} \phi^{vt} 
-ia \hat\epsilon^{\a\b}\chi^{vt}\right)V_{0n}^{ts}\label{VVmn}
\ee
where, by definition, $V_{0n}^{rs}=V_{n0}^{sr}$, and 
\be
\phi= \left(\matrix{1& -1/2& -1/2\cr 
                    -1/2& 1& -1/2\cr
                     -1/2 &-1/2 &1}\right)\label{phi}
\ee
while the matrix $\chi$ has been defined above (\ref{chi}). These two matrices
satisfy the algebra
\be
\chi^2 = - 2\phi,\quad\quad \phi\chi=\chi\phi = \frac 32 \chi,\quad\quad 
\phi^2= \frac 32 \phi\label{chiphi}
\ee

To end this section we would like to notice that the above results
can be easily extended to the case in which the transverse directions 
are more than two (i.e. the 24--th and 25--th ones) and even. The canonical
form of the transverse $B$ field is
\be
B_{\a\b} = \left(\matrix{\matrix{0&B_1\cr -B_1&0\cr} & 0& \ldots\cr
0& \matrix{0&B_2\cr -B_2&0\cr}&\ldots\cr
\ldots& \ldots& \ldots\cr}\right) \label{genB}
\ee 
It is not hard to see that each couple of conjugate transverse 
directions under this decomposition, can be treated in a completely
independent way. The result is that each couple of directions
$(26-i,25-i)$, corresponding to the eigenvalue $B_i$, will be 
characterized by the same formulas (\ref{VV00}, \ref{VV0n}, \ref{VVmn})
above with $B$ replaced by $B_i$.

\section{Properties of the new coefficients}

In this section we derive the properties of the coefficients 
$\V_{MN}^{\a\b,rs}$ which are essential for the later developments. 
These properties are parallel to those enjoyed by the ordinary 
coefficients, \cite{GJ1,GJ2,KP,RSZ2}. 

Let us quote first two straightforward properties of 
$\V_{MN}^{\a\b,rs}$:
\begin{itemize}
\item (i) they are symmetric under the simultaneous exchange of 
all the three couples of 
indices;
\item (ii) they are endowed with the property of cyclicity in the 
$r,s$ indices, i.e. $\V^{rs}= \V^{r+1,s+1}$, where $r,s=4$ is 
identified with $r,s=1$ and we have dropped the other indices.
\end{itemize}
The first property is immediate. The second can also be proven 
directly from eqs.(\ref{VVmn}). However, since it will be an easy 
consequence of eq.(\ref{basic}) below, we pass immediately to the 
derivation of the latter.

To this end we need the following representation of the coefficients 
$V_{0n}^{rs}$, derived from \cite{GJ1}:
\be
V_{0n}^{rs}= \left\{\matrix {Z_{n}\, \chi^{rs},\cr
-\frac 2{\sqrt 3}Z_{n}\, \phi^{rs},}\right. 
\quad\quad \matrix{ n\quad{\rm odd}\cr
n\quad{\rm even}}
\label{VZ}
\ee
where 
\be
Z_{n}= \sqrt{\frac 2{3n}}\, B_0A_n\label{Z}
\ee
The numbers $B_0$ and $A_n$ were defined in ref.\cite{GJ1}. Notice that,
since we have assumed $Z_{n}^{rs}=Z_{n}^{sr}$, we must have, 
by definition,
$V_{0n}^{rs}=V_{n0}^{rs}$ for $n$ even and 
$V_{0n}^{rs}=-V_{n0}^{rs}$ for $n$ odd.
Finally, for convenience, we introduce $Z_{0}=\sqrt \frac b3$.

Substituting (\ref{VZ}) into eqs.(\ref{VVmn}) and using (\ref{chiphi}), 
we obtain
\be
\V_{NM}^{\a\b,rs}= \left\{ \matrix{\V_{NM}^{\a\b,rs}(\infty) - 
\frac {6A^{-1}}{4a^2+3} 
K_\infty^{\a\b,rs} Z_{N}Z_{M},& \quad\quad N+M \quad{\rm even}\cr
 \V_{NM}^{\a\b,rs}(\infty) + 
\frac {\sqrt 3 A^{-1}}{4a^2+3} H_\infty^{\a\b,rs}
(-1)^NZ_{N}Z_{M}, &\quad\quad N+M \quad{\rm odd}\cr}\right.
\label{VVHKinf}
\ee
In these equations
\be
&&K_\infty^{\a\b,rs}= 
G^{\a\b}\phi^{rs}- ia  \hat\epsilon^{\a\b}\chi^{rs}\label{K}\\
&&H_\infty^{\a\b,rs}= 3G^{\a\b}\chi^{rs}+ 
4ia \hat\epsilon^{\a\b}\phi^{rs}\label{H}
\ee
and
$\V_{NM}^{\a\b,rs}(\infty)$ is  
\be
&&\V_{00}^{\a\b,rs}(\infty)= G^{\a\b} \delta^{rs}\0\\
&&\V_{0m}^{\a\b,rs}(\infty)= 0\label{VVinf}\\
&&\V_{nm}^{\a\b,rs}(\infty)= G^{\a\b}V_{nm}^{rs}\0
\ee
The coefficients $V_{nm}^{rs}$ are the same as in ref.\cite{RSZ2} for 
$n,m\geq  1$.

We can also express the $\V_{NM}^{\a\b,rs}$ in the following way
\be
\V_{NM}^{\a\b,rs}= \left\{\matrix{ \V_{NM}^{\a\b,rs}(0) 
+ \frac {6A^{-1}}{4a^2+3} 
K_0^{\a\b,rs} Z_{N}Z_{M}, &\quad\quad N+M \quad{\rm even}\cr
\V_{NM}^{\a\b,rs}(0) + 
\frac {\sqrt 3 A^{-1}}{4a^2+3} H_0^{\a\b,rs}
(-1)^NZ_{N}Z_{M},& \quad\quad N+M \quad{\rm odd}\cr}\right.\label{VVHK0}
\ee
where
\be
&&K_0^{\a\b,rs}=\frac 43 a^2 G^{\a\b}\phi^{rs}+ 
ia  \hat\epsilon^{\a\b}\chi^{rs}
\label{K0}\\
&&H_0^{\a\b,rs}= -4 a^2 G^{\a\b}\chi^{rs}+ 
4ia \hat\epsilon^{\a\b}\phi^{rs}\label{H0}
\ee
and $\V_{NM}^{\a\b,rs}(0) = G^{\a\b}V_{NM}^{'rs}$ are the values 
taken by 
$\V_{NM}^{\a\b,rs}$ for $B=0$. As expected, the symbols 
$V_{NM}^{'rs}$
are the same as the coefficients $V_{nm}^{'rs}(b)$ with $n,m\geq 0$, 
used in \cite{RSZ2}.  

Next we introduce the third root of unity $\omega= e^{i\frac {2\pi}3}$ 
and notice
that
\be
\phi^{rs}= \frac 12 (\omega^{r-s}+\omega^{s-r}),\quad\quad
\chi^{rs}= \frac i{\sqrt 3}(\omega^{r-s}-\omega^{s-r}),\label{omega}
\ee
Inserting these relations into (\ref{VVHKinf},\ref{VVHK0}) and 
rearranging the terms 
we find the basic relation
\be
\V_{NM}^{\a\b,rs}= \frac 13 \left(C'_{NM}G^{\a\b}+ 
\omega^{s-r}\U_{NM}^{\a\b} +\omega^{r-s} \bar \U_{NM}^{\a\b}\right)
\label{basic}
\ee
where
\be
\U_{NM}^{\a\b}= \left\{\matrix{G^{\a\b} \U_{NM}(\infty) + 
R^{\a\b} Z_{N}Z_{M}, &\quad\quad N+M\quad {\rm even}\cr
G^{\a\b} \U_{NM}(\infty) + iR^{\a\b}(-1)^N Z_{N}Z_{M}, 
&\quad\quad N+M \quad{\rm odd}\cr}\right.\label{UUinf}
\ee
Moreover
\be
\bar \U^{\a\b} = (\U^{\b\a})^*\label{barop}
\ee
where $^*$ denotes complex conjugation. In (\ref{basic}) $C'_{NM}= 
(-1)^N \delta_{NM}$
and
\be
R^{\a\b}= \frac {6A^{-1}}{4a^2+3}\left(-\frac 32 G^{\a\b}+\sqrt 3 a 
\hat\epsilon^{\a\b} \right)\label{R}
\ee
Moreover
\be
&&\U_{00}^{\a\b}(\infty)= G^{\a\b},\quad\quad \U_{0n}^{\a\b}=0\0\\
&&\U_{nm}^{\a\b}(\infty)= G^{\a\b}U_{nm}\label{Uinf}
\ee
In the last equation $U_{nm}$ coincides with the same symbol used in \cite{RSZ2}
(see eq.(B.15) in that reference).

Alternatively one can split $\U$ into the $B=0$ part and the 
rest. Then
\be
\U_{NM}^{\a\b} = \left\{\matrix{ G^{\a\b} \U_{NM}(0) + 
T^{\a\b} Z_{N}Z_{M}, &\quad\quad N+M\quad {\rm even}\cr
G^{\a\b} \U_{NM}(0) + iT^{\a\b}(-1)^N Z_{N}Z_{M}, &\quad\quad N+M
\quad {\rm odd}\cr}\right.\label{UU0}
\ee
where
\be
T^{\a\b}= \frac {12A^{-1}}{4a^2+3}\left(a^2 G^{\a\b}+\frac{\sqrt 3}2 a 
\hat\epsilon^{\a\b} \right)\label{T}
\ee
and $\U_{NM}^{\a\b}= G^{\a\b}  U'_{NM}$. The coefficients $U'_{nm},U'_{0n},U'_{00}$
are the same as in ref.\cite{RSZ2} (see eq.(B.19) therein).

Let us discuss the properties of $\U$. Since 
\be
(\U_{NM}^{\a\b})^*= \left\{\matrix{\U_{NM}^{\a\b},& 
\quad\quad  N+M \quad {\rm even}\cr
 -\U_{NM}^{\a\b},& \quad\quad N+M \quad {\rm odd}\cr}\right.\0
\ee
it is easy to prove the following properties (where we use the 
matrix notation for
the indices $N,M$)
\be
(\U^{\a\b})^* = C'\U^{\a\b}C'\label{UU*}
\ee
and
\be 
(\U^{\a\b})^\dagger = (\U^{\a\b})^{*T}= (C'\U^{\a\b}C')^T = \U^{\a\b}\label{Udag}
\ee
Finally, if tilde denotes transposition in the indices $\a,\b$, it is possible to
prove that (the proof is rather technical and deferred to Appendix A)
\be
(\U \tilde \U)_{NM}^{\a\b}= (\tilde\U \U)_{NM}^{\a\b}=               
G^{\a\b}\delta_{NM} + \left(RG +G\tilde R + \frac 23 AR\tilde R\right) Z_{N}Z_{M}
\label{UUtilde}
\ee
Now, remembering that 
$\hat\epsilon^{\a\gamma}\hat\epsilon_{\gamma}{}^\b= -G^{\a\b}\,$, 
it is elementary to prove that
\be
RG +G\tilde R + \frac 23 AR\tilde R=0\label{zero}
\ee 
Therefore, finally,
\be
(\U \tilde \U)_{NM}^{\a\b}= (\tilde\U \U)_{NM}^{\a\b}=\,G^{\a\b}\delta_{NM}
\label{UU=1}
\ee

Eqs.(\ref{UU*},\,\ref{Udag},\,\ref{UU=1}) are the generalization of the analogous ones
in \cite{GJ1,GJ2,KP,RSZ2}. Using in particular (\ref{UU=1}), it is easy to prove
that
\be
[C'\V^{rs},C'\V^{r's'}] =0.\label{CVCV}
\ee
This follows from
\be  
9[C'\V^{rs},C'\V^{r's'}] = \omega^{s-r+r'-s'} (C'\U\tilde \U C'- \tilde \U\U)
+ \omega^{s-r+s'-r'}(\tilde \U\U - C'\U \tilde \U C')\0
\ee
and from eq.(\ref{UU=1}). In the two previous equations matrix 
multiplication is understood both in the indices $M,N$ and $\a,\b$.
In the same sense, on the wake of \cite{KP,RSZ2}, we can also write 
down the following identities
\be
&&C'\V^{12}C'\V^{21}=C'\V^{21}C'\V^{12}=
 (C'\V^{11})^2- C'\V^{11}\label{CV1}\\
&&(C'\V^{12})^3+(C'\V^{21})^3 = 2(C'\V^{11})^3-
3(C'\V^{11})^2 +G\label{CV2}
\ee
which will be needed in the next section.

Notice however that, unlike refs.\cite{GJ1,GJ2,KP,RSZ2}, we have
\be
C'\V^{rs} = \tilde \V^{sr}C' \label{difference}
\ee

\section{The squeezed state solution}

A squeezed state in the present context is written as
\be
|S\rangle = |S_\perp\rangle\,\otimes\,|S_\|\rangle\label{decomp}
\ee
where $|S_\|\rangle$ has the ordinary form, see \cite{KP,RSZ2}, and is 
treated in the usual way, while
\be
\langle S_\perp| \,&=&\, {\cal N}^2 \,\langle \tilde 0|\,{\rm exp}
\left(-\frac 12 \sum_{M,N\geq 0}
a_M^{\a}\,\tilde\ES_{\a\b,MN}\, a_N^{\b}\right)
\label{squeezed1}\\
|S_\perp\rangle \,&=&\, {\cal N}^2 \,{\rm exp}\left(-\frac 12 \sum_{M,N\geq 0}
a_M^{\a\dagger}\,\ES_{\a\b,MN}\, a_N^{\b\dagger}\right)
|\tilde 0\rangle
\label{squeezed2}
\ee
where $|\tilde 0\rangle = |\Omega_{b,\theta}\rangle\otimes |0\rangle$.
Here we have written down both bra and ket in order to stress the 
difference with \cite{GJ1,GJ2,KP,RSZ2}, which stems from the 
fact that, in view of (\ref{difference}), we assume
 $C'\ES^{\a\b}C'=(\ES^{\a\b})^*= \ES^{\b\a}$.
The $*$ product of  two such states, labeled $_1$ and $_2$, is carried out
in the same way as in the ordinary case, see again \cite{KP,RSZ2}. Therefore
we limit ourselves to writing down the result
\be
|S'_\perp\rangle =|S_{1,\perp}\rangle\,*\,|S_{2,\perp}\rangle= 
\frac{K_2\,({\cal N}_1{\cal N}_2)^2}{{\rm DET}({\bf I}- 
\Sigma{\cal V})^{1/2}}\, \,
{\rm exp}\left(-\frac 12 \sum_{M,N\geq 0}
a_M^{\a\dagger}\ES'_{\a\b,{MN}}a_N^{\b\dagger}\right)
|\tilde 0\rangle \label{squeez}
\ee
where, in matrix notation which includes both the indices $N,M$ and $\a,\b$,
\be
\ES'= \V^{11} +(\V^{12},\V^{21})({\bf I}- 
\Sigma{\cal V})^{-1}\Sigma 
\left(\matrix{\V^{21}\cr \V^{12}}\right)\label{SS'}
\ee
In RHS of these equations 
\be
\Sigma= \left(\matrix{\tilde\ES_1&0\cr 0& \tilde\ES_2}\right),\quad\quad\quad
{\cal V} = \left(\matrix{\V^{11}&\V^{12}\cr \V^{21}&\V^{22}}\right),
\label{SigmaV}
\ee
and ${\bf I}^{\a,rs}_{\b,MN}= \delta^\a_\b\,\delta_{MN}\,\delta^{rs}$,
$r,s= 1,2$. ${\rm DET}$ is the determinant with respect to all indices.
To reach the form (\ref{SS'}) one has to use cyclicity of $\V^{rs}$ (property
(ii) above). The expression of $\ES'$ is in fact a series, therefore some
kind of condition on the coefficients $\ES_i$ must be satisfied in order for it to
make sense. The squeezed states $\ES$ satisfying this condition form
a subalgebra of the algebra defined by the $*$ product.

Let us now discuss the squeezed state solution of the equation 
$|\Psi\rangle * |\Psi\rangle =|\Psi\rangle$ in the matter sector. In order for this to be satisfied
with the above states $|S\rangle$, we must first impose
\be
\ES_1=\ES_2=\ES'\equiv\ES\0
\ee
and then suitably normalize the resulting state.
Then (\ref{SS'}) becomes an equation for $\ES$, i.e.
\be
\tilde \ES= \V^{11} +(\V^{12},\V^{21})({\bf I}- \Sigma{\cal V})^{-1}\Sigma 
\left(\matrix{\V^{21}\cr \V^{12}}\right)\label{SS}
\ee
where $\Sigma,{\cal V}$ are the same as above with $\ES_1=\ES_2=\ES$.
Eq.(\ref{SS}) has an obvious (formal) solution by iteration. However 
in ref. \cite{KP} it was shown that it is possible to obtain the solution in
compact form by `abelianizing' the problem. 
Notwithstanding the differences with that case, it is possible
to reproduce the same trick on eq.(\ref{SS}), thanks to (\ref{CVCV}).
One denotes $C'\V^{rs}$ by $\X^{rs}$ and $C'\ES$ by $\T$, and
assumes that $[\X^{rs},\T]=0$ (of course this has to be checked 
{\it a posteriori}). Notice however that we cannot assume that
$C'$ commutes with $\ES$, but we assume that $C'\ES = \tilde \ES C'$. 
By multiplying (\ref{SS}) from the left by $C'$ we get:
\be
 \T= \X^{11} +(\X^{12},\X^{21})({\bf I}- \Sigma{\cal V})^{-1}
\left(\matrix{\T\X^{21}\cr \T\X^{12}}\right)\label{TT}
\ee
For instance $\tilde\ES \V^{12}= \tilde\ES C' C' \V^{12}= \T \X^{12}$, etc.
In the same way,
\be
({\bf I}- \Sigma{\cal V})^{-1}= 
\left( \matrix{{\one}-\T\X^{11}& -\T\X^{12}\cr
-\T \X^{21}& {\one}- \T\X^{11}}\right)^{-1}\0
\ee 
where $\one^{\a}_{\b,MN}= \delta^\a_\b\,\delta_{MN}$. Now 
all the entries are commuting matrices, so the inverse can
be calculated straight away.

From now on everything is the same as in \cite{KP,RSZ2}, therefore we 
limit ourselves to a quick exposition.
Using (\ref{CV1}) and (\ref{CV2}), one arrives at an 
equation only in terms of $\T$ and $\X
\equiv \X^{11}$:
\be
(\T-\one)(\X\T^2- (\one+\X)\T +\X)=0\label{fineq}
\ee
This gives two solutions:
\be
&&\T =\one\label{sol1}\\
&&\T = \frac 1{2\X}\left( \one +\X - \sqrt{(\one + 3\X)(\one-\X)}\right)
\label{sol2}
\ee
The third solution, with a + sign in front of the square root, is not acceptable,
as explained in \cite{RSZ2}. In both cases we see that the solution commutes with
$\X^{rs}$. Naturally we are talking about solutions of the abelianized
eq.(\ref{TT}). The true solution we are looking for is, in both cases,
$\ES=C'\T$.

As for (\ref{sol1}), it is easy to see that it leads to the identity state.
Therefore, from now on we will consider (\ref{sol2}) alone.

Now, let us deal with the normalization of $|S_\perp\rangle$. 
Imposing $|S_\perp\rangle\, *\,|S_\perp\rangle = |S_\perp\rangle$ we find
\be
{\cal N}^2 = K_2^{-1} \,{\rm DET}\, ({\bf I} - \Sigma {\cal V})^{1/2}\0
\ee
Replacing in it the solution one finds
\be
 {\rm DET} ({\bf I} - \Sigma {\cal V}) = {\rm Det}\,
 \left( (\one -\X)(\one + \T)\right)
\label{det}
\ee
${\rm Det}$ denotes the determinant with respect to the indices 
$\a,\b,M,N$.
Using this equation and (\ref{K2}), and borrowing from \cite{RSZ2}
the expression for $|S_\|\,\rangle$, one finally gets for the 23--dimensional
tachyonic lump:
\be
|S\rangle \!&=&\! \left\{{\rm det}(1-X)^{1/2}{\rm det} (1+T)^{1/2}\right\}^{24}
{\rm exp}\left(-\frac 12 \eta_{\bar \mu\bar \nu}\sum_{m,n\geq 1} 
a_m^{\bar \mu\dagger}S_{mn}a_n^{\bar \nu\dagger}\right)|0\rangle
\otimes\label{fullsol}\\
&& \frac {A^2 (3+4a^2)}{\sqrt{2 \pi b^3}({\rm Det}G)^{1/4}} 
\left( {\rm Det}(\one -\X)^{1/2}{\rm Det}(\one + \T)^{1/2}\right)
{\rm exp}\left(-\frac 12 \sum_{M,N\geq 0}
a_M^{\a\dagger}\ES_{\a\b,MN}a_N^{\b\dagger}\right)|\tilde 0 \rangle,\0
\ee
where $\ES= C'\T $ and $\T$ is given by (\ref{sol2}). 
The quantities in the first line are defined in ref.\cite{RSZ2} with $\bar\mu,\bar\nu=0,
\ldots 23$ denoting the parallel directions to the lump.

The value of the action corresponding to (\ref{fullsol}) is easily calculated
\be
{\cal S}_\ES\!&=&\! {\EuScript K} \frac {V^{(24)}}{(2\pi)^{24}}
\left\{{\rm det}(1-X)^{3/4}{\rm det} (1+3X)^{1/4}\right\}^{24}\0\\ 
&&\cdot\, \frac {A^4 (3+4a^2)^2}{{2 \pi b^3}({\rm Det}G)^{1/2}} \,\,
{\rm Det} (\one -\X)^{3/4}{\rm Det}(\one + 3\X)^{1/4}\label{actionS}
\ee
where $V^{(24)}$ is the volume along the parallel directions and ${\EuScript K}$ is the
constant of eq.(\ref{action2}). 

Finally, let $\mathfrak e$ denote the energy per unit volume, 
which coincides with the brane tension when $B=0$. Then one can 
compute the ratio of the D23--brane energy density 
${\mathfrak e}_{23}$ to the D25-brane 
energy density ${\mathfrak e}_{25}$ ;
\be
\frac {{\mathfrak e}_{23}}{ {\mathfrak e}_{25}} &\!=\!&
\frac {(2\pi)^2} {({\rm Det}G)^{1/4}}\cdot 
{\EuScript{R}} \label {ratio1} \\
\R &\!=\!&  
\frac {A^4 (3+4a^2)^2}{2 \pi b^3({\rm Det}G)^{1/4}}
\frac {{\rm Det}(\one -\X)^{3/4}{\rm Det}(\one + 3\X)^{1/4}}
{{\rm det}(1 -X)^{3/2}{\rm det}(1 + 3X)^{1/2}}\label{ratio2}
\ee

If the quantity $\R$ equals 1 (see the comment below on this issue), 
this equation is 
exactly what is expected for the ratio of a flat static D25--brane
action and a D23--brane action per unit volume in the presence 
of the $B$ field (\ref{B}). In fact the DBI Lagrangian for a flat 
static Dp--brane is, \cite{SW},
\be
{\cal L}_{DBI} = \frac 1{g_s (2\pi)^p} \sqrt { {\rm Det} (1+ 2\pi
B)}\label{DBI}
\ee
where $g_s$ is the closed string coupling.
Substituting (\ref{B}) and taking the ratio the claim follows.

To end this section let us briefly discuss the generalization of the
above results to lower dimensional lumps. As remarked at the end of 
section 2, every couple of transverse directions corresponding to
an eigenvalue $B_i$ of the field $B$ can be treated in the same way 
as the 24--th and 25--th directions. One has simply to replace in the
above formulas $B$ with $B_i$. The derivation of the above formulas
for the case of $25-2i$ dimensional lumps is straightforward.

\section{A comment}

Switching on a constant $B$ field on VSFT does not obstruct 
the possibility to find exact results. On the contrary, we have
found that (matter) squeezed states representing tachyonic lumps are
still solutions of the equations of motion, and that we can give
compact explicit formulas for these solutions, much like in the 
$B=0$ case. These are still interpretable as (lower dimensional)
D--branes. We have seen that the expected ratios for their tensions 
are reproduced, modulo the hypothesis that in (\ref{ratio1},
\ref{ratio2}), the ratio $\R$ be equal to 1. Let us discuss 
this briefly. 

Let us recall that, as $B\to 0$ ($a\to 0$),
\be
\X_{MN}\to X'_{MN}\label{0lim}
\ee
where $X'$ is the matrix used in \cite{RSZ2} to define lower 
dimensional tachyonic lumps. It was shown there, numerically, that
\be
\frac {9A^4}{2 \pi b^3}
\frac {{\rm det}(1 -X')^{3/2}{\rm det}(1 + 3X')^{1/2}}
{{\rm det}(1 -X)^{3/2}{\rm det}(1 + 3X)^{1/2}}\label{ratioXX'}
\ee
is actually 1, and that this result does not seem to depend on 
the values taken by the parameter $b$.
Now, (\ref{ratioXX'}) is exactly the limit of (\ref{ratio2}) when 
$a\to 0$.  Therefore, looking at the structure of $\X$, it seems
reasonable to assume that (\ref{ratio2}) is also 1, at least
as long as $a$ is not too large. In the limit $a\to \infty$ 
the exponent in the second line of (\ref{fullsol}) becomes 
ill--defined and some kind of rescaling becomes necessary.
However, before going deeply into this issue and the previos 
assumption, we believe that a better knowledge
of the eigenvalues of $\X$ is needed. This requires an
ad hoc analysis, which will be done elsewhere along the lines of
\cite{RSZ5}.

\vskip0.5cm

{\bf Note Added}. After we had completed this work, there appeared
a new paper by K.Okuyama, \cite{oku3}, who, using in particular
the results of \cite{RSZ5} and \cite{HM}, was able to analytically 
prove that (\ref{ratioXX'}) is exactly 1. Using the same kind
of arguments we have been able to prove analytically that
in fact the ratio $\R$ in (\ref{ratio2}) is also 1. The long 
details of this calculation will appear elsewhere, \cite{BMS}, 
together with new results on VSFT in the presence of a $B$ field.

\acknowledgments

We would like to thank Martin Schnabl for calling our attention
on ref.\cite{oku3}.
This research was supported by the Italian MIUR 
under the program ``Teoria dei Campi, Superstringhe e Gravit\`a".

\appendix

\section{Derivation of $(\U \tilde \U)_{NM}^{\a\b}$}

In this Appendix we derive eq.(\ref{UUtilde}). This can be done 
starting both from the representation (\ref{UUinf}) and from 
(\ref{UU0}). In the first case we need
the following identities taken from the Appendix B of \cite{RSZ2}. 
\be
\sum_{n\geq 1} W_n U_{nm} 
=W_m,\quad\quad\sum_{n\geq 1}W^*_nW_n= 2V_{00}\label{Wn}
\ee
The numbers $W_n$ are defined via the equation
\be
V_{0n}^{rs}= \frac 13 (\omega^{s-r}W_n + \omega^{r-s} W_n^*)\label{defWn}
\ee
On the other hand we have
\be
&&V_{0n}^{rs}= \frac {i}{\sqrt 3}(\omega^{r-s} - \omega^{s-r})Z_{n},
\quad\quad n \quad {\rm odd}\0\\
&&V_{0n}^{rs}= -\frac {1}{\sqrt 3}(\omega^{r-s} + \omega^{s-r})Z_{n},
\quad\quad n \quad {\rm even}\label{V0nZ0n}
\ee
This allows us to identify $W_n$ and $Z_{n}$ as follows:
\be
&&W_n = -i \sqrt 3 Z_{n}, \quad\quad n \quad {\rm odd} \0\\
&&W_n = -\sqrt 3 Z_{n}, \quad\quad n \quad {\rm even} \label{WZ}
\ee
In particular, from the second equation in (\ref{Wn}), we get
\be
\sum_{n\geq 1} Z_{n}^2= \frac 23 V_{00}\label{Z0n2}
\ee

Next one has to consider $(\U \tilde \U)_{NM}$ case by case according 
to the various possibilities for $N,M$. As a sample, let us consider 
$N=n$ odd and $M=m$ odd. Then
\be
(\U \tilde \U)_{nm}= \U_{n0}\tilde\U_{0m} + \sum_{k\,{\rm odd}} \U_{nk}\tilde
\U_{km} +   \sum_{k\,{\rm even}} \U_{nk}\tilde \U_{km} \0
\ee
Now we replace on the RHS the values extracted from eq.(\ref{UUinf}). 
After rearranging the terms we get
\be
(\U \tilde \U)_{nm}&= &G \delta_{nm} + \frac b3 R \tilde R Z_{n}Z_{m}+
R\tilde R Z_nZ_m \sum_{k\geq 1}Z_k^2\0\\
&&- \frac i{\sqrt 3} G\tilde R \sum_{k\geq 1} U_{nk}W_k^* Z_m 
+ \frac i{\sqrt 3}R G Z_n \sum_{k\geq 1} W_kU_{km}\0\\
&= &G \delta_{nm} \left(RG +G\tilde R + \frac 23 (V_{00}+ 
\frac b2)R \tilde R\right)
Z_nZ_m\label{finid}
\ee
where use has been made of (\ref{Wn}) and (\ref{Z0n2}). In the same way all
other cases of the identity (\ref{UUtilde}) can be proved.

Alternatively one can prove (\ref{UUtilde}) by means of the representation 
(\ref{UU0}).
The procedure is the same, but the matrix involved is $U'$ instead of $U$. For
this reason we need, instead of the second eq.(\ref{Wn}), the identity
\be
\sum_{n\geq 1} W_n U'_{nm} = \frac {\frac b2 - V_{00}}{\frac b2 + V_{00}} W_m
\ee

\end{document}